\documentclass{Interspeech2024}
\usepackage{url}
\interspeechcameraready

\title{Accent conversion using discrete units with parallel data synthesized from controllable accented TTS}

\name[affiliation={1}]{Tuan-Nam}{Nguyen}
\name[affiliation={1}]{Ngoc-Quan}{Pham}
\name[affiliation={1}]{Alexander}{Waibel}

\address{
  $^1$ Karlsruhe Institute of Technology , Germany}
\email{tuan.nguyen@kit.edu, ngoc.pham@kit.edu, alexander.waibel@kit.edu}


\begin{document}
\maketitle
\begin{abstract}
The goal of accent conversion (AC) is to convert speech accents while preserving content and speaker identity. Previous methods either required reference utterances during inference, did not preserve speaker identity well, or used one-to-one systems that could only be trained for each non-native accent. This paper presents a promising AC model that can convert many accents into native to overcome these issues. Our approach utilizes discrete units, derived from clustering self-supervised representations of native speech, as an intermediary target for accent conversion. Leveraging multi-speaker text-to-speech synthesis, it transforms these discrete representations back into native speech while retaining the speaker identity. Additionally, we develop an efficient data augmentation method to train the system without demanding a lot of non-native resources. Our system is proved to improve non-native speaker fluency, sound like a native accent, and preserve original speaker identity well.     
\end{abstract}

\section{Introduction}

Accents or the struggle in understanding accents can be a language barrier between different English speaking groups. Deep learning models have the capability to address this issue by effectively converting accents while retaining the speakers’ identity.

Conventional accent conversion \cite{8462258,guanlong2019-interspeech}methods rely on reference utterances in the target accent during synthesis, which restricts their applicability. Such limitation arises from the difficulty in obtaining reference utterances with identical linguistic content yet in a different accent. Similar to the most recent research in accent conversion, this research concentrates on reference-free AC, which can convert accent without a reference utterance during inference. Approaches to reference-free AC can be categorized into two main architectures: non-autoregressive and autoregressive. 

Non-autoregressive AC \cite{jia2023zeroshot,jin2022voicepreserving} only entails non-parallel data that comprise diverse but unpaired information and are widely accessible. Making use of such data can require decomposing speech into distinct independent features such as speaker identity, content, prosody and accent. Although the method enables synchronization between the input and output audios, non-autoregressive models may encounter challenges due to the lack of fluency in source non-native speakers. It handles accent conversion without altering the duration of the input audio, hence hardly improves the fluency quality \cite{jia2023zeroshot}.

In contrast, autoregressive accent conversion is primarily based on seq2seq model and uses parallel data for training. Parallel data consist of utterances from the same speaker in different accents, which can be challenging to obtain because of its scarcity. To address the issue, one approach \cite{AC_NamNguyen} employs data augmentation techniques by utilizing voice conversion to generate parallel data with similar voices but different accents. Nevertheless, the method was not experimented with unseen speakers (zero-shot condition) and are one-to-one directed systems that must be trained independently for each non-native accent. Another approach is from \cite{quamer22_interspeech,9477581} which does not utilize any data augmentation techniques. Instead, it trains a model to "translate" non-native bottleneck features derived from phonetic posteriorgrams into equivalent native bottleneck ones. This method effectively corrects pronunciation errors in non-native utterances. Then they convert these native bottleneck features into corresponding mel-spectrogram and eventually into waveform audio using a neural vocoder. Their training data come from accented speech recordings instead of synthesized speech data, rendering difficulty in making use of the training data that represent the same text content in multiple accents.

\begin{figure}[!ht]
\begin{center}
\includegraphics[scale=0.21]{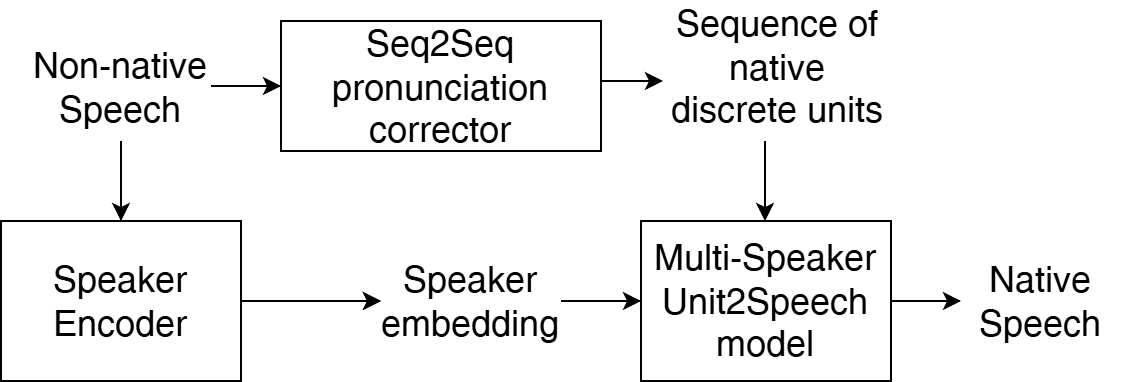} 
\caption{The workflow of the proposed system}
\label{fig.1}
\vspace{-0.5cm}
\end{center}
\end{figure}

Our research developed an autogressive, reference-free, zero-shot, and many-to-one directional AC system that can be trained in a low non-native resource condition and improve the fluency for non-native speech. Additionally, how the capability of generating diverse accented speech with the same text content can help training parallel AC models is also investigated, and this data augmentation strategy set our work apart from other works. Figure \ref{fig.1} illustrates our method, which cascades a seq2seq pronunciation corrector (PC)  model and a native multi-speaker unit-to-speech (U2S) model. The PC model converts plenty of non-native accented speech into discrete representation units, which are obtained from the clustering of self-supervised speech representations on native speech. With these self-supervised discrete units \cite{lakhotia2021generative} being able to separating the content of speech from the speaker identity, converting these discrete units back to speech can be done with the original speaker identity. The second model, multi-speaker U2S, utilizes the speaker embedding generated by speaker encoder for this purpose. Within this study, the units derived from native speech are even more easily convertible back to native speech. Nevertheless, our evaluation is grounded in real data. The contributions of this work are as follows:


\begin{itemize}
    \item The proposed system can transform speech inputs of varied accents into native while preserving the speaker's voice and content. Additionally, it can improve the fluency of non-native speakers. 
    \item A substantial amount of parallel training data may be required for a seq2seq PC. Therefore, we devised a data augmentation technique to generate more training data under a limited non-native resources condition. Parallel synthetic training data is introduced to effectively train the Voice Conversion (VC) system \cite{9689675}. We believe it can also be used to train any AC systems in a parallel fashion without an initial demand for such kind of data.
    \item Our proposal is to pioneer the utilization of discrete units in the training of a transformer-based model for AC. Inspired by the success of self-supervised pretrained models on Speech Recognition \cite{pham2022adaptive} and Speech-to-Speech Translation \cite{DBLP:journals/corr/abs-2107-05604}, we show that pretrained encoder-decoder can also boost training effectiveness more than random weight initialization.

\end{itemize}

\section{Methodology}
\begin{figure*}[t]
  \centering
  \includegraphics[scale=0.17]{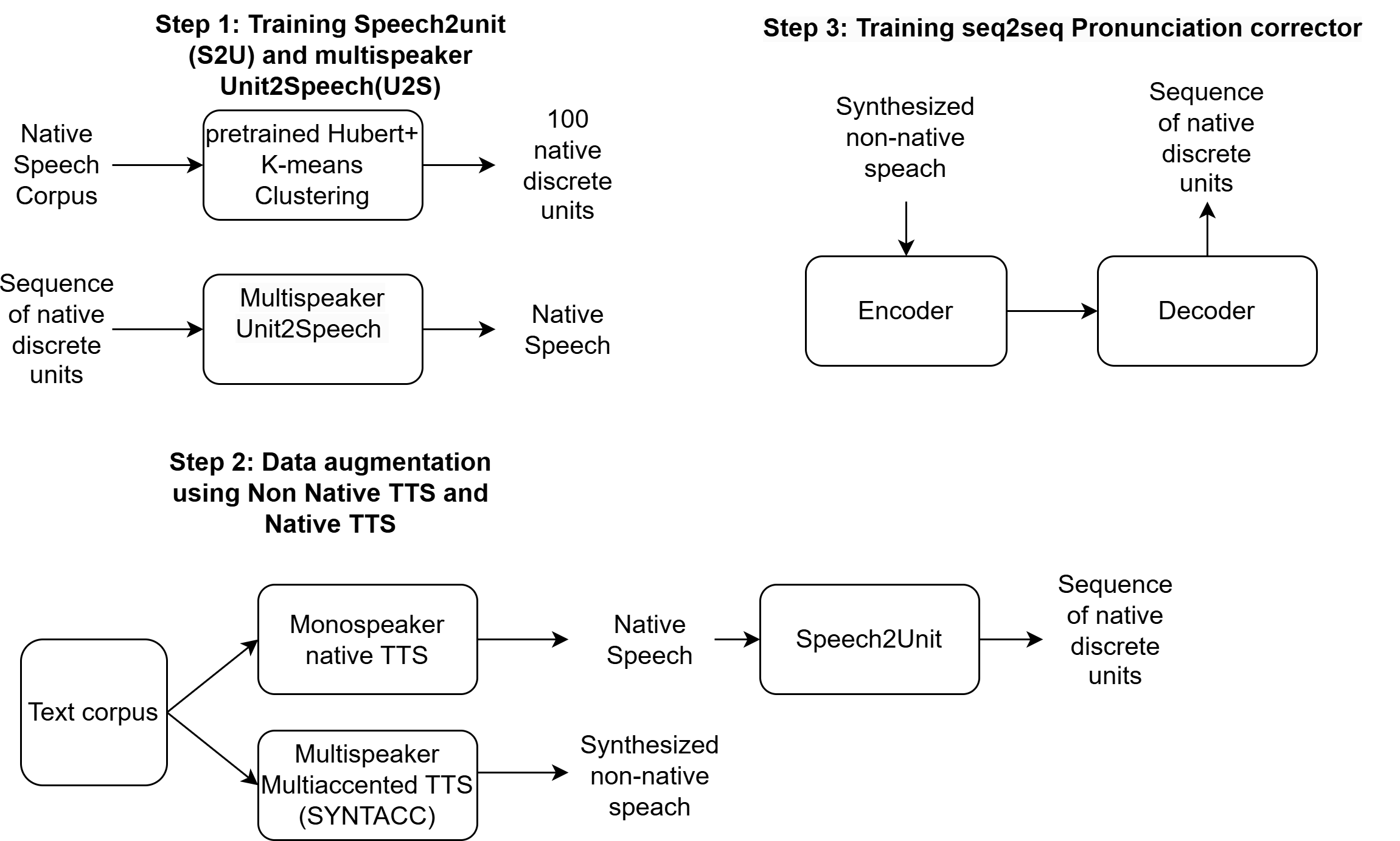}
  \caption{Three steps of training system}
  \label{fig:Accent TTS architecture}
\end{figure*}

Figure \ref{fig:Accent TTS architecture} illustrates the three steps to achieve the first unit-based AC system. First, we train S2U and U2S using native speech corpus. Then we train multi-speaker multi-accented TTS and Monospeaker native TTS to create parallel training data. Finally, we use the synthesized paralel training data for training the seq2seq PC.
\subsection{Speech2Unit model and Multi-speaker Unit2Speech model}\label{Speech2Unit model and Native Unit2Speech model}

Following the best setting in \cite{lakhotia2021generative}, we utilize a HuBERT model~\cite{Hubert} to encode native speech corpus into continuous representations at every 20-ms frame, and to learn K-means with K = 100 on these representations from the sixth layer \footnote{Empirically the middle layer produces the codes with the best quality}. To generate discrete unit sequences, we quantize these representations by mapping each one to its nearest cluster based on the Euclidean distance. We remove consecutive duplicate units to construct a reduced unit sequence representing native speech..

YourTTS \cite{yourTTS} is a zero shot multi-speaker TTS that can achieve good voice similarity for unseen speakers, hence we decide to use YourTTS architecture for U2S. We treat the discrete units extracted from native speech as text input, and train YourTTS separately on a native corpus with a large number of speakers. Finally, our U2S can convert native discrete units with speaker embedding back to any original native speech.

\subsection{Data augmentation using Multi-accented TTS and Native TTS}
In order to train the discrete unit generative model, one can question how we can provide enough data consisting of multiple accents. Here approach is to start with a good base TTS model, and then enhance it with a multi-accented adaptation step to control the TTS to generate into any accent. Following the work in SYNTACC (SYNThesizing speech with multiple ACCents) ~\cite{SYNCTACC}, with YourTTS as the base model, we employed weight factorization~\cite{DBLP:journals/corr/abs-2105-03010} to achieve the multi-accented model. Each weight matrix in the TTS model is factorized into a shared component and an accent-specific component. While the former is initialized by the pretrained conventional multi-speaker TTS model, the latter is simplified as rank-1 matrices to not only minimize the memory cost for each extra accent, but also to encourage the shared component to hold as much information as possible while each accent only needs a few parameters to control.

 This adaptation stage is effective in fine-tuning the SYNTACC in the absence of a large demand for non-native speech data, while also retaining the ability to synthesize unseen voices of the original multi-speaker TTS model. Finally, the SYNTACC model can synthesize speech in not only multiple voices, but also varied accents. 


We train YourTTS on the native speaker corpus to generate output audios. The seq2seq PC model's outputs are discrete units, hence these training sequences should be consistent. To ensure consistent voice, style, and accent in synthetic output, we train YourTTS with audio from a single native speaker. Then applying S2U on synthesized output audios to creates discrete native unit sequences.


After training the Native YourTTS and SYNTACC models, we use them to create training data for the PC. First, we require a large text corpus. Then, for each sentence in this text corpus, we generate corresponding input and output audio. We produce input audio for each sentence using SYNTACC with a random speaker and accent. We can additionally modify the synthesized audio by randomizing the duration noise scale and the inference noise scale of SYNTACC to make the seq2seq PC model more robust on a variety of input sounds. In contrast to the SYNTACC model, we fix these noise scales when doing inference native YourTTS, then use S2U to construct a sequence of discrete native units for output audio. Finally, in the following section, the input non-native audios and matching sequence of discrete  units can be utilized to train the seq2seq PC.

\subsection{Training the pronunciation corrector}
Our proposed PC is a Transformer based sequence-to-sequence model with a speech encoder and a discrete unit decoder. In this work, we explore both pretrained encoder and decoder.
\subsubsection{Pretrained Encoder: Wav2vec 2.0 and  HuBERT}
The discrete units are derived from the representations of the pretrained HuBERT, we regard this pre-trained model as a viable choice for initializing encoder weights. Wav2vec 2.0  \cite{Wav2vec} is also a self-supervised frameworks to learn speech representations from unlabeled audio data. They both use a multi-layer convolution neural network to encode the audio followed by a Transformer-based context encoder to build the contextualized representations. In this work, we use Wav2vec 2.0 and HuBERT with relative attention \cite{pham20_interspeech} for better performance.
\subsubsection{Pretrained Decoder: MBart50}
BART was originally proposed for denoising autoencoder over text using Transformer. The network is tasked to reconstruct a sentence at the decoder given a noisy version at the encoder side. MBart50 \cite{MBart} and its extension MBart50 took the BART training scheme and applied to 50 languages. Our discrete units were obtained from English data, therefore we can consider these discrete units to be a new language that has some similarities to English and can benefit from pretrained MBart50 decoder. In our case, we replace embedding layer of MBart50 and treat the discrete units as text output and traing wav2vec-MBart and HuBERT-MBart on our synthetic parallel data.
\section{Experiments}
\subsection{Data and training description}

Speech2Unit and Unit2Speech models are obtained from LJSpeech corpus \cite{ljspeech17} and LibriTTS-R (the sound quality improved version of the LibriTTS corpus \cite{LibriTTS,librittsr}corpus, which has more than 2300 speakers) ,respectively. The LJSpeech corpus contains 13,100 audio samples of a single native female voice reading sentences. We follow the section~\ref{Speech2Unit model and Native Unit2Speech model}  to learn S2U for English speech. To obtain a multi-speaker U2S, we train a multi-speaker YourTTS with discrete units from S2U associated with speaker embedding as input and target speech as output. We use pretrained speaker encoder \cite{wan2020generalized} to generate speaker embedding.

We use audio data from the LJSpeech corpus and L2-Arctic \cite{zhao2018l2arctic} to train the YourTTS and fine-tune the SYNTACC for data augmentation respectively. L2-Arctic corpora include Hindi accent, Mandarin accent, Vietnamese accent, as well as Korean accent, Spanish accent and Arabic accent. It has a total of 24 speakers, each of whom is featured in their own audio recordings of the same 1152 sentences. Totally, we get around 3 hours of each non-native accent, which can be considered as a low-resource condition. We fine-tune SYNTACC and train YourTTS from scratch using the settings from the original papers. We utilize CommonVoice corpus's transcripts to generate parallel training data as stated in Section 3.1. The Coqui-TTS \footnote{\url{https://github.com/coqui-ai/TTS}} is used for training all TTS and U2S models. To train S2U model, we use the source code and setting from fairseq \footnote{\url{https://github.com/facebookresearch/fairseq/tree/main/examples/textless_nlp/gslm/speech2unit}}.

In our research, we also investigate how data augmentation can affect the performance of the PC. In details, we devise two ways for synthesizing 1 million utterances. We can select one million sentences from a text corpus and generate one non-native audio recording of each sentence with a random accent and speaker, which we call non-overlapped sentence strategy. The second strategy involves selecting 166 thousand sentences from a text corpus and creating non-native 6 audio files per sentence, each sentence with all 6 accents and 6 random speakers, which we call overlapped sentence strategy. In both circumstances, we generate one native audio per sentence as output audio. The second way, we believe, could help PC learn how to address varied accents more successfully.  Before generating synthetic data, we divided the sentences into train and validation sets in the ratio 1000:1. The test data is our in-house  data, which contains approximately 1000 sentences (3 hours) recorded by Chinese, Indian, Arabic and Vietnamese speakers. These speakers have not been recorded in training data, so we can consider as zero-shot condition.  

To train the PC, we utilized the wav2vec 2.0 and HuBERT pretrained with the Large configuration and hidden size of 1024. The MBart50 decoder employs the same hidden size. We employed an effective batch size of roughly 1 hour of audio with a gradient accumulation technique for each update during training with a linear decay learning rate schedule, starting from~$0.001$. The model takes 20k updates to converge (about 5 to 6 hours with a single GPU RTX A6000 with 48GB of RAM). We employ beam search with beam size 8 for inference. The Hugging Face framework \footnote{\url{https://github.com/huggingface}}is used for training PC. 

\subsection{Evaluation metrics}

\textbf{Test Perplexity}
For the PC, we estimate the perplexity on our in-house test data. Perplexity is a measure of how efficiently a model predicts the next unit in a sequence of units. In our case, it also indicates how well the PC learns the patterns of native speech and decodes them in the discrete units. For each sentence in test set, we use the native YourTTS to synthesize a native audio with the same content, followed by Speech2Unit to generate a ground-truth sequence of units. These ground-truth sequences can be used to estimate the perplexity of the PC on the test data. Furthermore, to assess how data augmentation and pretrained encoder-decoder affect performance, we need to compare different data augmentation strategies and different weight initialization methods (with and without pretrained model). Then the best PC with the lowest test perplexity is chosen when evaluating the subjective metrics of the whole system.

\textbf{Accentedness test, Fluency test, Speaker Similarity Mean Opinion Score (Sim-MOS) and Mean Opinion Score (MOS).} To evaluate the performance of the whole system, we use three subjective metrics. We pick 50 random sentences in tets set for evaluation. For each test sentence, three kind of tests are conducted by 20 American participants who listen to the provided audios and evaluate their overall quality on a 5-point scale: 1-bad, 2-poor, 3-fair, 4-good, 5-excellent. In the Accentedness test and Fluency test, the participants give a score for the degree to which the synthesized audios sound like a native speech and how much they speak fluently, respectively. For the Sim-MOS test, they rate the similarity between the voice timbre of the output audios and the original audios of target speakers on a scale of 5 as above. Finally, we compute the mean with 95{\%} confidence interval for all subjective metrics.

We have 3 most recent baseline systems. The first one \cite{quamer22_interspeech} is also the seq2seq-based system,  they convert a non-native speech to mel-spectrogram of native speech, then use an external vocoder to convert back to waveform. The next two baselines \cite{jia2023zeroshot},\cite{jin2022voicepreserving} are non-autoregressive systems with disentanglement network to disentangle accent attribute from original speech. Due to the unavailability of the source codes, we compared our system's outputs to their best audio examples. Sample evaluation audios are available at a github repository \footnote{\url{https://accentconversion.github.io/}}. 

\subsection{Results}

Table \ref{tab:result1} illustrates that the sentence overlapping setting significantly outperforms the non-overlapping setting in terms of text perplexity under all weight initialization conditions. It can indicate that the data which has many accents, many speaker for each sentence is very suitable for our AC problem, we believe it can help the PC learn how to distinguish multiple accents better. In term of weight initialization, the combination of Wav2vec encoder and MBart Decoder has best PPL. Consequently, this combination, along with the sentence overlapping setting, is selected for the subjective test.
\begin{table}

	\setlength{\tabcolsep}{4pt}
	
	\centering 
 
	\begin{tabular}{lc}
	\cline{1-2}
         Models &  PPL\\ 
       \cline{1-2}

        No pretrained  \\
         \quad +  non-overlapped sentence  & 3.63  \\
         \quad  +  overlapped sentence &   2.45 \\
        \cline{1-2}
        Wav2vec encoder + no pretrained decoder \\
         \quad  +   non-overlapped sentence& 3.21  \\
         \quad + overlapped sentence  & 2.25\\
        \cline{1-2}
        HuBERT encoder + no pretrained decoder \\
         \quad  +   non-overlapped sentence& 3.28  \\
         \quad + overlapped sentence  & 2.32\\
        \cline{1-2}
        Wav2vec encoder  + MBart decoder \\
         \quad  + no sentence overlapped & 3.11  \\
         \quad + sentence overlapped & \textbf{2.16} \\
        \cline{1-2}
        HuBERT encoder + MBart decoder \\
         \quad  +   non-overlapped sentence& 3.23  \\
         \quad + overlapped sentence  & 2.24\\
        \cline{1-2}
	\end{tabular}
\vspace{0.5cm}
\caption{Test perplexity}
\label{tab:result1}
\end{table}

\begin{table}[ht]

	\setlength{\tabcolsep}{4pt}
	\centering

	\begin{tabular}{lccc}
		 \cline{1-4}
        \textbf{Models} & Accentness & Sim MOS & Fluency \\ 
     \cline{1-4}
         Input  & $2.12\pm0.05$ &   & $3.31\pm0.03$   \\
        \cline{1-4}
         Proposed   \\
          \quad{In-house test}  & $\mathbf{4.46}\pm0.12$ & $3.89\pm0.09$ & $\mathbf{4.55}\pm0.07$ \\
        
           \quad{Public test}  & $\mathbf{4.42}\pm0.11$ & $3.95\pm0.07$& $\mathbf{4.51} \pm0.06$ \\
         \cline{1-4}
         Baseline-1 & $3.67\pm0.11$ & $3.61\pm0.06$ & $3.83\pm0.08$ \\
         Baseline-2 & $3.65\pm0.09$ & $3.92\pm0.05$ & $3.94\pm0.09$ \\
         Baseline-3 & $3.93\pm0.10$ & $\mathbf{4.23}\pm0.10$ & $3.84\pm0.10$ \\
        
		 \cline{1-4}
    
	\end{tabular}
	\vspace{0.5cm}
 \caption{Subjective metrics}
\label{tab:result2}
\end{table}
In the table \ref{tab:result2}, for the audio quality, our model outperform Baseline-1 in all metrics. It can indicates that native discrete units is better representation than mel-spectrogram in our AC problem. To compare with other two baseline, our model is better in term of accentedness and fluency, but slight worse than the Baseline-3 in term of speaker similarity. Such observation implies that the our model might have changed the audios to a greater extent compared to the other baseline, which actually produces a more native output yet makes the participants can find the speaker identity altered more. These two non-autoregressive baseline can keep the duration of the input audio while resulting in little change in fluency. They convert accent without modifying the input audios’ duration, allowing the input and output audios can be synchronized, making them ideal for applications such as dubbing a video with accents. Our system can significantly improve fluency, making it more suited for language understanding applications. Our model performs comparably well in both in-house and public test sets.

\section{Conclusion}\label{sec:reference}

In this paper, we have shown that a decent controllable accented TTS could provide a convenient way to generate huge amount of parallel training data for AC. We also believe that this data augmentation strategy can help generate effectively more accented training data for Speech Recognition and  Translation. It is also described how a pretrained encoder decoder with native discrete units can contribute to the training of a many-to-one directional AC system. Experimental results show that the proposed method is able to convert unseen speakers’ utterances into the native accent with better fluency and accent. Further study will focus on improving the ability to keep speaker identity.

\section{Acknowledgement}
This research was supported in part by a grant from Zoom Video Communications, Inc.  The authors gratefully acknowledge the support. Part of this work was supported by funding from the pilot program Core Informatics of the Helmholtz Association (HGF).

\bibliographystyle{IEEEtran}
\bibliography{mybib}

\end{document}